\documentclass[aps, prb, twocolumn, amsmath,amssymb]{revtex4}
\usepackage{graphicx}
\usepackage{bm}

\begin{document}

\title{Order parameter symmetry and mode coupling effects at dirty superconducting quantum phase transitions}

\author{Rastko Sknepnek}

\author{Thomas Vojta}
\affiliation{Department of Physics, University of Missouri - Rolla, Rolla, MO, 65409, USA}

\author{Rajesh Narayanan}
\affiliation{Institut f{\" u}r Nanotechnologie, Forschungszentrum Karlsruhe, 76021 Karlsruhe,
Germany}


\begin{abstract}
We derive an order-parameter field theory for a quantum phase transition between a disordered
metal and an exotic (non-$s$-wave) superconductor. Mode coupling effects between the order
parameter and other fermionic soft modes lead to an effective long-range interaction between the
anomalous density fluctuations which is reflected in singularities in the free energy functional.
However, this long-range interaction is not strong enough to suppress disorder fluctuations. The
asymptotic critical region is characterized by run-away flow to large disorder. For weak coupling,
this asymptotic region is very narrow. It is preempted by a wide crossover regime with mean-field
critical behavior and, in the $p$-wave case, logarithmic corrections to scaling in all dimensions.

\end{abstract}

\maketitle

\section{Introduction}
\label{sec:Int} Quantum phase transitions are one of the most intriguing problems in today's
condensed matter physics.\cite{Hertz_76,Sondhi_97,Sachdev_99} In addition to being of
fundamental interest, they are believed to underlie a number of interesting low-temperature
phenomena, in particular various forms of exotic superconductivity.
\cite{exotic_sc,fm_sc_exp1,fm_sc_exp2}

In a seminal paper,\cite{Hertz_76} Hertz introduced a general scheme for the analysis of
quantum phase transitions in itinerant electronic systems. This scheme is based on the
Landau-Ginzburg-Wilson (LGW) approach of integrating out the fermionic degrees of freedom and
deriving a free energy functional in terms of the order parameter fluctuations only. However,
in recent years, it has become clear that for many transitions, there are problems with Hertz'
scheme because besides the order parameter fluctuations, which are soft (gapless) at the
critical point, there are additional fermionic soft modes in the system.  These additional soft
modes exist not only at the critical point but also in the bulk phases. They are related to
conservation laws and/or broken symmetries and constitute examples of generic scale
invariance.\cite{gsi} If the coupling between the order parameter and these additional soft
modes is sufficiently strong it generates an effective long-range interaction between the order
parameter fluctuations. This is reflected in a non-analytic wave-number dependence of the
vertices of the LGW theory of the corresponding
transition.\cite{kbv_clean_fm,millis_2d,chubukov_2d} Generically, such non-localities in the
LGW theory will lead to non-mean-field critical behavior of the quantum phase transition.

The precise influence of the mode-coupling effects on a quantum phase transition depends on the
structure of the additional soft modes and their coupling to the order parameter. For the clean
ferromagnetic transition, the mode-coupling effects can either lead to a fluctuation-induced
first order transition or to non-mean-field critical behavior.\cite{us_fm_clean,bk_fm_clean}
For dirty electrons, the transitions is generically of second order but with highly unusual
exponents.\cite{kb_dirty_fm,bk_dirty_q} Even stronger effects were found for the transition
between a dirty metal and a conventional ($s$-wave) superconductor.\cite{kb_scqpt_97} Here, the
mode-coupling effects lead to a critical point with exponential scaling, i.e., the correlation
length behaves as $\xi\sim e^{1/|t|}$, where $t$ is distance from the quantum critical point.
Based on general symmetries of itinerant electronic systems, it was recently shown
\cite{us_nonlocal} that homogeneous (${\bf q}=0$) \cite{afm} order parameters in the
particle-particle (Cooper) and spin-triplet particle-hole channels are generically affected by
mode coupling effects while order parameters in the particle-hole spin-singlet channel do allow
for a local LGW theory.

All of the above examples are quantum phase transitions with zero angular momentum order
parameters. The effect of mode-coupling on order-parameters with finite angular momentum are
much less understood. Herbut \cite{Herbut_00} studied the $d$-wave superconducting quantum
phase transition in two dimensions within Hertz' scheme (which is equivalent to Gorkov's
mean-field theory). He found that the typical Cooper channel (BCS) logarithmic singularities
are demoted to irrelevant terms by the $d$-wave symmetry. This raises the important general
question: How does a finite order parameter angular momentum influence the coupling between
order parameter and additional fermionic soft modes?

In this Paper, we study this question for quantum phase transitions between a metal and an
exotic (non-$s$-wave) superconductor in the presence of non-magnetic quenched disorder. These
transitions are of experimental importance since various superconducting states with p,d and
maybe higher symmetries have been observed recently, and their quantum phase transitions are
experimentally accessible \cite{exotic_sc,fm_sc_exp1,fm_sc_exp2} in principle. Experiments
performed on the weakly ferromagnetic compounds UGe$_2$ \cite{fm_sc_exp1} and ZrZn$_2$
\cite{fm_sc_exp2} revealed the existence of a superconducting phase within the ferromagnetic
phase at temperatures below 1K. It is believed\cite{fm_sc_exp1} that both superconductivity and
ferromagnetism arise from the same electrons. One possible mechanism for this type of
superconductivity is $p$-wave triplet pairing mediated by magnetic fluctuations in the vicinity
of a magnetic quantum critical point,\cite{FayAppel,us_fm_sc} although this is still not a
settled issue. The onset of the phenomenon has proven to be very sensitive to the presence of
non-magnetic disorder, making it observable only in highly pure samples. This fact also points
to a non-$s$-wave order parameter. In ZrZn$_2$ the superconducting quantum phase transition as
a function of disorder has already been observed.\cite{fm_sc_exp2}

Our results can be summarized as follows: Mode-coupling induced singularities exist for all
order parameter angular momenta $L$. However, with increasing $L$, they are more and more
suppressed (by a factor $|{\bf q}|^{2L}$.) As a result, the LGW theory of our superconducting
transition is equivalent to that of the itinerant antiferromagnetic
transition.\cite{Hertz_76,CardyBoyanovsky,bk_afm,us_rr} While the asymptotic critical behavior
of this theory is not understood because of runaway flow to large disorder, we also show that
for weak bare disorder the asymptotic region is exponentially narrow. It is preempted by a wide
crossover regime with mean-field critical behavior and (in the $p$-wave case) logarithmic
corrections to scaling. The paper is organized as follows.  In Sec.\ \ref{sec:Model}, we derive
the LGW free energy functional. In Sec.\ \ref{sec:rg}, we study the LGW theory by means of the
renormalization group and determine the critical behavior. In Section \ref{Sec:Conclusions}, we
analyze our findings from a mode-coupling point of view, and  discuss differences between
paramagnetic and ferromagnetic as well as  clean and dirty systems.

\section{Landau-Ginzburg-Wilson  theory}
\label{sec:Model}
\subsection{$p$-wave pairing case}
\label{sec:ModelA} In this section, we derive an effective LGW theory for the disorder-driven
quantum phase transition between a paramagnetic metal and a $p$-wave triplet superconductor.
Our starting point is a microscopic action for interacting electrons in $d>2$ dimensions and
subject to a static, single-particle random potential $V(\mathbf x)$. We assume a Gaussian
distributed potential with $[V(\mathbf x_1)V(\mathbf x_2)]_{dis} = W\delta(\mathbf x_1- \mathbf
x_2)$, with $W$ being measure of disorder strength. The partition function $Z$ can be written
as a functional integral over Grassmann variables $\psi,\bar \psi$:
\begin{equation}
Z=\int D[\bar \psi, \psi] e^{S[\bar \psi, \psi]}~.
\end{equation}
We decompose the action $S=S_p+S_0$ into the $p$-wave interaction part $S_p$ and a reference system
$S_0$ which comprises the single-particle part, the random potential and $S_{int}$ (the
interaction in all channels other than the $p$-wave channel):
\begin{eqnarray}
S_0 & = & \int dx \sum_\sigma \bar \psi_\sigma(x) [\partial_\tau  + \frac{\nabla^2}{2m} + \mu]
\psi_\sigma(x) + \\ & + & \int dx \sum_\sigma \bar \psi_\sigma(x) V(\mathbf x) \psi_\sigma(x) +
S_{int},\label{eq:S_0}
\end{eqnarray}
\begin{eqnarray}
S_{p}  &=&
\sum_{\{\sigma\}}\frac{\Gamma^{\{\sigma\}}_t}{2} \int dx \bar {\mathbf n}_{\sigma\sigma'}(x)
\cdot \mathbf n_{\sigma_1\sigma'_1}(x)~.\label{eq:S_int}
\end{eqnarray}
We use a ($d+1$)-vector notation, with $x = (\mathbf x, \tau)$, $k = (\mathbf k, \Omega)$,
$\int dx = \int_V d^dx \int_0^\beta d\tau$ and $\sum_k = \sum_{\mathbf k}T\sum_{\Omega}$,
$\mathbf x$ is a real space coordinate, $\tau$ imaginary time, $\mathbf k$ momentum vector and
$\Omega$ Matsubara frequency. ${\mathbf n}_{\sigma\sigma'}(x)$ is the $p$-wave anomalous
density whose Fourier transform in terms of the fermion fields is given by
\begin{equation}
 \mathbf{n}_{\sigma\sigma'}({q})= \sum_k {\mathbf{\hat{e}}_\mathbf{k}} ~\psi_\sigma(k+q/2) \psi_{\sigma'}(k-q/2),
\end{equation}
with  $\mathbf{\hat{e}}_\mathbf{k} = \mathbf k/|\mathbf k|$, $\sigma$, $\sigma'$ being spin
indices and $\cdot$ denoting scalar product in the orbital space. Due to the Pauli principle the
spin state of the Cooper pair has to be a triplet, i.\ e.\
$\sigma\sigma'\in\{(\uparrow\uparrow),(\downarrow\downarrow),1/\sqrt{2}(\uparrow\downarrow+\downarrow\uparrow)\}$.
Which combination of the three possible triplet components is actually realized depends on the
system under consideration. The reference ensemble $S_0$ describes interacting electrons in the
presence of non-magnetic quenched disorder and no bare interaction in the $p$-wave Cooper channel.
(A non-vanishing interaction in this channel will be generated in perturbation theory.) $S_0$ thus
describes a general system of disordered interacting electrons with the only restriction being
that it must not undergo a phase transition in the parameter region we are interested in.

A standard procedure \cite{Hertz_76} is used to derive a LGW order-parameter field theory. We
decouple the interaction term using a Hubbard-Stratonovich transformation
\cite{Hubbard_59,Stratonovich_57} introducing a complex field
$\mathbf{\Delta}_{\sigma\sigma'}(x)$ which plays the role of an order parameter. Quenched
disorder treated using the replica trick,\cite{replica} and fermionic degrees are  integrated
out, leading to an expression for the critical part of the partition function in terms of
order-parameter only:
\begin{equation}
Z = \int {\rm D}[\mathbf{\Delta}] ~e^{-\Phi[\mathbf{\Delta]}}~.
\end{equation}
Since our emphasis is on the mode-coupling effects, and in order to avoid unnecessary
complications in notation, we restrict our analysis to a certain spin component
($\sigma\sigma'=(\uparrow\uparrow)$) of the order parameter. The LGW free energy $\Phi[\mathbf
\Delta]$ is expanded in powers of the order parameter
$\mathbf{\Delta}\equiv\mathbf{\Delta}_{\uparrow\uparrow} $. Up to quartic order it reads:
\begin{eqnarray}
\label{eqn:LGW_first} \Phi[\mathbf \Delta] &=& \sum_{q,\alpha} \bar {\mathbf
\Delta}^\alpha(q)(1-\Gamma_t\chi^{(2)}(q))\mathbf \Delta^\alpha(q) \nonumber \\ &-&\sum_{q_1 \dots
q_3\atop\alpha,\beta} \Gamma_t^2\chi^{(4)}_{\alpha\beta}(q_1,q_2,q_3) \bar {\mathbf
\Delta}^\alpha(q_1){\mathbf \Delta}^\alpha(q_2) \nonumber \\
 &\times&\bar {\mathbf \Delta}^\beta(q_3){\mathbf \Delta}^\beta(q_1+q_3-q_2) + O(\Delta^6),
\end{eqnarray}
where $\Gamma_t^{\uparrow\uparrow}\equiv\Gamma_t$. Here $\alpha,\beta$ are replica indexes. The
coefficients of the LGW functional are determined by the 2-point and 4-point anomalous density
correlation functions of the reference system $S_0$ which can be written as $\chi^{(2)}=\langle \bar
{n}^\alpha n^\alpha \rangle_{0}$ and $\chi^{(4)}_{\alpha\beta}=\langle \bar {n}^\alpha
\bar{n}^\beta {n}^\beta n^\alpha\rangle_{0}$ (with the spin and component indices suppressed).

\subsection{Anomalous density correlation functions in the $p$-wave channel}
\label{Sec:Model_correl_fun}

In this subsection we use diagrammatic perturbation theory in disorder and interactions to
calculate  $\chi^{(2)}$ and $\chi^{(4)}$ of the reference ensemble $S_0$, focusing on the
behavior for weak disorder.  Thus, we neglect all diagrams with crossed impurity lines, i.e.,
all weak localization effects, which become important only at higher impurity concentrations.
This is justified for the ferromagnetic superconductors \cite{fm_sc_exp1,fm_sc_exp2} where the
superconducting quantum phase transition occurs at very small disorder.

We start our analysis by examining the 2-point correlation function, $\chi^{(2)}=\langle \bar
{n}^\alpha n^\alpha \rangle_{0}$. The leading contributions are obtained from the diagrams shown
in Fig.\ \ref{Fig:chi2} (details of the calculation of these diagrams are given in Appendix
\ref{appx:bubble}).
\begin{figure}
\includegraphics[width=\columnwidth]{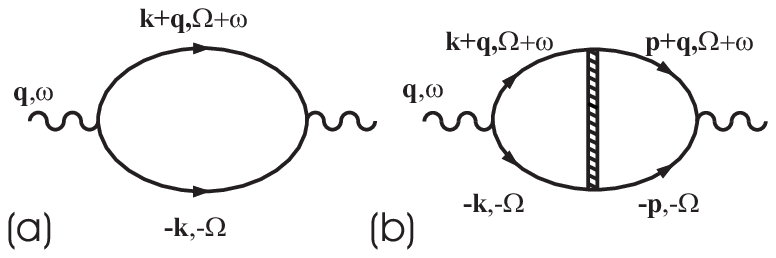}
\caption{Contributions to the leading terms of the Gaussian part of the LGW functional. (a)
provides a constant ($\sim N_F$) and the frequency dependence, $|\omega|\tau$, while (b) gives the
leading momentum dependence, $|{\bf q}|^{2} \log(1/|{\bf q}|)$.} \label{Fig:chi2}
\end{figure}
Here, the external vertices represent anomalous $p$-wave densities, the solid lines are fermionic
propagators in Born approximation,
\begin{equation}
\label{eq:Green_fun} G_{\sigma}^{-1}(\mathbf k, \omega) = i\omega - \epsilon_{\mathbf k,\sigma}
+\mu + (i/2\tau)\textrm{sgn} (\omega),
\end{equation}
where $\epsilon_{\mathbf k, \sigma}$ is the dispersion relation and $\tau$ is the scattering time.
The double line represents a particle-particle impurity ladder:
\begin{equation}
\label{eq:ladder}
W_R(\mathbf q, \Omega, \omega)=W\left\{
\begin{array}{cl}
1 &\textrm{if $\Omega(\Omega+\omega)<0$}\\
\frac{1}{|2\Omega+\omega|\tau+\ell^2|{\mathbf q}|^2/d}&\textrm{if $\Omega(\Omega+\omega)>0$}
\end{array} \right.
\end{equation}
$\ell=k_F\tau/m$ is the elastic mean free path and $W = 1/(2\pi N_F \tau)$ with $N_F$ being the
density of states at the Fermi level. The calculation of the diagrams in Fig.\ \ref{Fig:chi2} for
$|{\bf q}| =\omega=0$ leads to $\chi^{(2)}= (N_F/3) \ln(2\epsilon_F\tau)$. The well-known
logarithmic Cooper channel (BCS) singularities are cut-off by the disorder, reflecting the
suppression of exotic superconductivity by non-magnetic scatterers in analogy to the suppression
of $s$-wave superconductivity by magnetic impurities.\cite{gorkov,larkin} [We note that, in
contrast, $s$-wave superconductivity is not influenced by weak non-magnetic scatterers, as is
signified by Anderson's theorem.\cite{anderson_th}] However, a closer investigation of diagram
\ref{Fig:chi2}b) for finite $|{\bf q}|$ reveals that a non-analyticity of the form $|{\bf q}|^2
\ln (1/|{\bf q}|)$ survives. Thus, the $p$-wave symmetry has demoted the BCS singularity to
quadratic order in $|{\bf q}|$ because each of the renormalized external vertices picks up an
extra power of $|{\bf q}|$.\cite{adachi}

In addition to the BCS logarithms  $\chi^{(2)}$ contains non-analyticities similar to that in
the itinerant ferromagnet. They are caused by the leading corrections to scaling at the dirty
Fermi liquid fixed point \cite{bk_review} and can be viewed as particle-particle analogs of the
well known Altshuler-Aronov corrections to density of states and conductivity.\cite{Altshuler}
For $s$-wave order parameters these singularities (which only arise for interacting electrons)
are of the form $|{\bf q}|^{d-2}$. \cite{bk_review} For $p$-wave order parameters, an
inspection of the corresponding contributions (details see Appendix \ref{appx:int}) reveals
that they are suppressed by a factor $|{\bf q}|^2$ by the same mechanism as the BCS logarithms.
[Note that an analogous suppression occurs in the particle-hole channel, as can be seen from a
power counting analysis of the Altshuler-Aronov correction to the conductivity: $\delta \sigma
= \delta\langle j j\rangle/\omega \sim \omega^{(d-2)/2} \sim |{\bf q}|^{d-2}$. Thus, the
correction to the current-current correlation function
 $\langle j j\rangle$, which is proportional to the $p$-wave density, behaves as $\delta\langle j
j\rangle \sim |{\bf q}|^d$. This means it has picked up an additional factor $|{\bf q}|^2$ compared to the zero
angular momentum channel.]

As a result, we find that the leading singularities in $\chi^{(2)}$ are the Cooper channel
logarithms, and the leading behavior of $\chi^{(2)}$ for $(|{\bf q}|,\omega)\to 0$ is given by:
\begin{eqnarray}
\chi_{ij}^{(2)}({\bf q},\omega) &=& \frac{N_F}{3}[\ln(2\epsilon_F\tau) - |\omega|\tau
                    -\frac{\ell^2|{\mathbf q}|^2}{10}]\delta_{ij}+ \nonumber \\
&&+\frac{N_F}{3}q_iq_j \ell^2[-\frac{1}{5}+\frac{1}{3}\ln(\frac{\ell^2|{\mathbf q}|^2}{3})].
\label{eqn:chi_2}
\end{eqnarray}
where $i,j$ are the order parameter component indices. The anisotropic $\bf q$-dependence in the
last term reflects the spatial anisotropy of the $p$-wave order parameter.

\begin{figure}
\begin{center}
\includegraphics[scale=0.9]{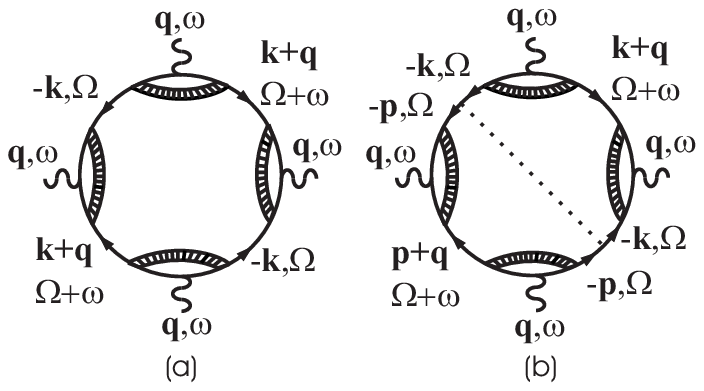}
\end{center}
\caption{Leading singular contributions to $\chi^{(4)}_{diag}$. After expansion in small $q$,
the leading order terms of (a) and (b) cancel (see Appendix \ref{appx:football}).
 } \label{Fig:chi4d}
\end{figure}

We now turn our attention  to the 4-point correlation function, $\chi^{(4)}$which can be split
into a replica-diagonal part and a replica-off-diagonal part, $\chi^{(4)}_{\alpha\beta} =
\delta_{\alpha\beta}\chi_{diag}^{(4)} + \chi_{off}^{(4)}$. A detailed discussion of our
calculation is given in the Appendix \ref{appx:football}. We find that the leading contributions
to $\chi_{diag}^{(4)}$ in the long-wavelength, low-frequency limit are coming from the diagrams
shown in Fig.\ \ref{Fig:chi4d}.
While each of the two diagrams  individually diverges for $({\bf q},\omega)\to0$, their leading
singularities cancel, and the remaining contribution is finite and proportional to $N_F \tau^2$.
\begin{figure}
\begin{center}
\includegraphics[scale=1.2]{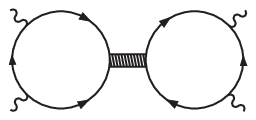}%
\end{center}
\caption{Leading contribution to the $\chi^{(4)}_{off}$ (see Appendix \ref{appx:football}).}
\label{Fig:goggle_bare}
\end{figure}
The leading contribution to the replica-off-diagonal part of $\chi^{(4)}$  is produced by the
diagram shown on the Fig.\ \ref{Fig:goggle_bare}. Thus, we finally obtain:
\begin{eqnarray}
\chi_{diag}^{(4)} &=& A_d~N_F \tau^2 ~F_d(\mathbf q_i, \omega_i)~,\label{eq:chi_4diag}\\
\chi_{off}^{(4)} &=& A_o ~ ({N_F^2}/{k_F^3})~F_o(\mathbf q_i, \omega_i)~,\label{eq:chi_4off}
\end{eqnarray}
where $F_d$ and $F_o$ are dimensionless functions with values between 0 and 1  and $A_d$ and $A_o$
are dimensionless prefactors of order one.

The results  (\ref{eq:chi_4diag}) and (\ref{eq:chi_4off}) have been obtained from low order
perturbation theory. Within perturbation theory, it is non-trivial to prove the leading
non-analyticity to all orders. Therefore we follow the guidance of the corresponding results for
the $s$-wave case which have been rigorously established using $Q$-field-theory and
renormalization group arguments.\cite{bk_review} (Corresponding rigorous results for finite
angular momentum modes do not yet exist.) Indeed, simply correcting the $s$-wave results from Ref.
\onlinecite{kb_scqpt_97} for the $q$-dependence of the renormalized vertex leads to $\chi^{2n} \sim |{\bf q}|^{4-2n}$ in agreement with
(\ref{eq:chi_4diag}) and (\ref{eq:chi_4off}). Note that the singularity becomes stronger in the
higher order anomalous density correlation functions, in agreement with general mode-coupling
arguments.\cite{us_nonlocal,divvert}

Inserting eqs.\ (\ref{eqn:chi_2})-(\ref{eq:chi_4off}) into Eq.\ (\ref{eqn:LGW_first}) we obtain
the LGW functional up to quartic order in $\Delta$,
\begin{eqnarray}
 \Phi&=&\sum_{q,\alpha\atop i} {\Delta}^{\ast\alpha i}(q)
                      (t+|\omega|+c_{2}|{\mathbf q}|^2) \Delta^{\alpha i}(q) \nonumber \\
    &+& \sum_{q,\alpha\atop ij}{\Delta}^{\ast\alpha i}(q)  c_{l} q_i q_j [-\frac{1}{5}+\frac{1}{3}\ln(\frac{1}{\ell^2|{\mathbf q}|^2})] \Delta^{\alpha j}(q)
    \nonumber\\
  &+&U\sum_\alpha\int d\mathbf r d\tau~|\Delta^\alpha(\mathbf r,\tau)|^4 \nonumber\\
 &-&V\sum_{\alpha\beta}\int d\mathbf r d\tau d\tau'~|\Delta^\alpha(\mathbf r,\tau)|^2|\Delta^\beta(\mathbf r,\tau')|^2~.
\label{eqn:LGW_second}
\end{eqnarray}
Here we have scaled the order parameter with $({\Gamma_t N_F \tau})^{1/2}$ and replaced the
quartic coefficients by numbers which is sufficient for power counting purposes. The
coefficients are: $t\sim1/\tau\left[1/\tilde\Gamma-\ln({2\epsilon_F\tau})\right]$, $c_{2}\sim
c_{l}\sim\ell^2/\tau$, $U\sim 1/N_F$, and $V\sim 1/\left(k_F^3 \tau^2\right)$, with
$\tilde\Gamma = \Gamma_t N_F$ being a dimensionless measure of the interaction strength. The
parameter $t$ represents the distance from the quantum critical point. Generically, $U>0$,
which leads to a second order transition.\cite{chubukov_1st} This completes the derivation of
the LGW theory.

\subsection{Higher angular momentum channels}
\label{Sec:ModelB}

In this subsection we generalize the findings to pairings in higher angular momentum ($L$)
channels. For angular momentum $L>0$, the renormalized anomalous density vertex is proportional
to $|{\bf q}|^L$. The leading non-analyticity in the static anomalous density susceptibility
$\chi^{(2)}_{L}({\mathbf q})$  is given by the BCS logarithms in diagram \ref{Fig:chi2}b. They
take the form
\begin{equation}
 \delta \chi^{(2)}_{L}({\mathbf q}) =\delta \langle \bar{n}_L^M({q}) n_L^M({q})\rangle \sim |{\mathbf q}|^{2L} \log(1/|{\mathbf q}|),
\label{eq:sing}
\end{equation}
i.e., they are suppressed by a factor $|{\mathbf q}|^{2L}$ compared to the $s$-wave case. Here
$n_L^M({ q}) = \sum_k Y_L^M({\mathbf{\hat{e}}_\mathbf{k}}) \psi_\sigma(k+q/2)
\psi_{\sigma'}(k-q/2)$ is a component of the anomalous density for angular momentum $L$. Note that
for $L>1$ the BCS logarithm is subleading compared to the regular $|{\bf q}|^2$ term coming from
diagram \ref{Fig:chi2}a) while in the $p$-wave case the BCS logarithm provides the leading
wave-number dependence in the LGW functional. The same mechanism also suppresses the interaction
induced mode-coupling singularities related to corrections to scaling at the dirty Fermi liquid
fixed point. An explicit calculation outlined in Appendix \ref{appx:int} shows that these mode
coupling singularities behave at most like $|{\mathbf q}|^{2L} |{\mathbf q}|^{d-2}$ ($d$ is the
spatial dimensionality). Therefore they are sub-leading compared to the BCS logarithms for all
$d>2$. We now turn to the 4-point anomalous density correlation function $\chi^{(4)}_L$. Because
of the momentum dependence of the renormalized anomalous density vertex, $\chi^{(4)}_L$ picks up
an extra power of $|{\bf q}|^{4L}$ compared to the $s$-wave case. More generally, any $2n$-point
anomalous density correlation function $\chi^{(2n)}$ picks up an extra power of $|{\bf q}|^{2nL}$
compared to the $s$-wave case, i.e. $\chi^{(2n)}\sim|{\bf q}|^{4+(2L-4)n}$. Therefore their
singular contributions are demoted to sub-leading order and do not play a role for the critical
behavior.

As a result, the leading terms in the LGW functional for $d$-wave and higher order parameter
angular momentum take the same form (\ref{eqn:LGW_second}) as in the $p$-wave case except for
the missing logarithmic wavenumber dependence in the Gaussian part.\cite{Herbut_00}

\section{Renormalization Group Analysis}
\label{sec:rg}

In this section, we analyze the  LGW theory (\ref{eqn:LGW_second}) by means of the
renormalization group. There is a Gaussian fixed point with mean-field static exponents
$\nu=1/2$, $\gamma=1$, $\eta=0$, and a dynamical exponent of $z=2$. In the $p$-wave case there
are logarithmic corrections to the mean-field behavior in all dimensions. In order to check the
stability of this Gaussian fixed point we study the importance of quantum and disorder
fluctuations. The scale dimensions of $U$ and $V$ at the Gaussian fixed point can be calculated
by power counting. We obtain $[U]=2-d$ and $[V]=4-d$. Thus, the conventional fluctuation term
(the $U$ term) is renormalization group irrelevant for $d>2$, but the disorder term (the
replica-off-diagonal quartic $V$ term) is relevant for $d<4$. In three dimensions the Gaussian
fixed point is unstable with respect to the disorder term, and thus the calculation of loops is
necessary to determine the asymptotic critical behavior. This includes the possibility of
replica-symmetry breaking in the replica-off-diagonal quartic term.

\begin{figure}
\begin{center}
\includegraphics[scale=0.6]{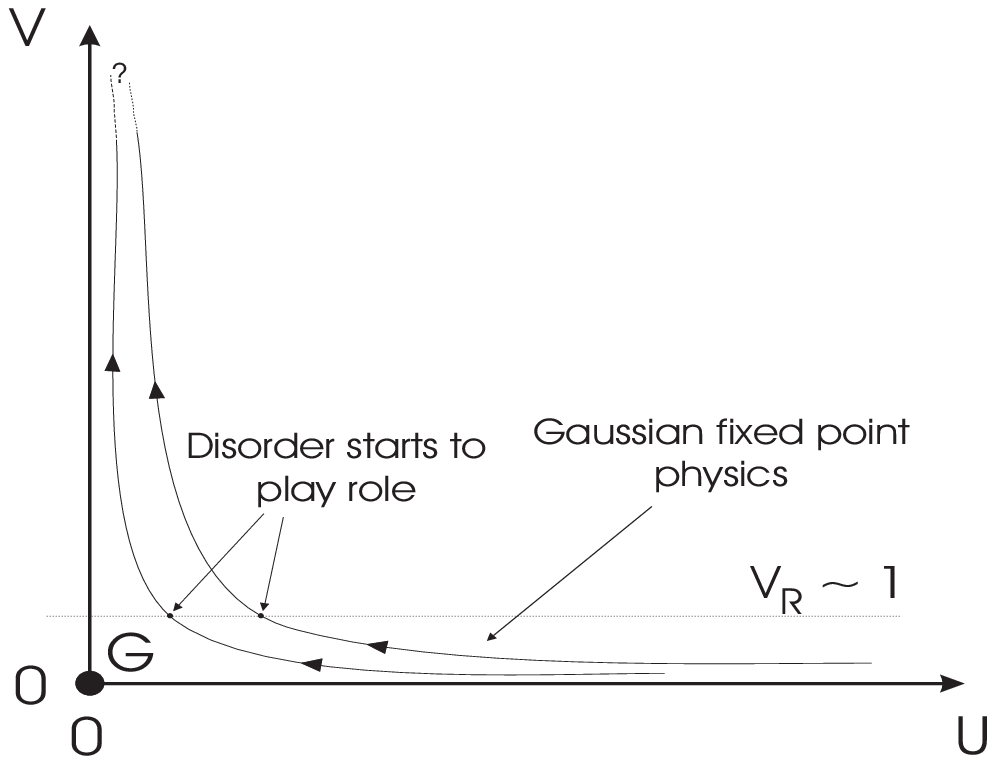}%
\end{center}
\caption{Schematic flow diagram on the critical surface. The Gaussian fixed point (G) is
unstable; the flow goes toward large disorder $V$. For weak bare disorder, the flow stays close
to the $U-$axis until it almost reaches the Gaussian fixed point before crossing over (black
dots) to the asymptotic destination. The dashed line separates the region described by the
Gaussian fixed point from the strong disorder region. } \label{Fig:rg_flow}
\end{figure}

Rather than carrying out this program explicitly, we use the analogy between our transition and
the disordered itinerant antiferromagnetic quantum phase transition to discuss the asymptotic
critical behavior: Except for the logarithmic corrections, the LGW theory, eq.\
(\ref{eqn:LGW_second}) is identical to that of a disordered itinerant antiferromagnet. This
transition has been investigated in great detail in recent years.
\cite{CardyBoyanovsky,bk_afm,us_rr} By taking into account rare disorder fluctuations it was found
that there is no critical fixed point in the perturbative region of parameter space, and the
asymptotic critical behavior is characterized by run-away flow toward large disorder (Fig.\
\ref{Fig:rg_flow}) rendering the perturbation expansion unjustified. The physical implications of
this runaway flow are not fully understood so far. Possible scenarios include a non-perturbative
fixed point with conventional (power-law) scaling, an infinite randomness fixed point (relative
magnitude of inhomogeneities increases without limit under coarse graining), resulting in
activated scaling, or a complete destruction of a sharp phase transition. Thus, from the analogy
with the quantum phase transition in itinerant antiferromagnets we conclude that the asymptotic
critical behavior of our theory is unconventional, and, at present, unknown.

However, in many relevant experimental systems the bare disorder is actually very small. Thus, one
may ask at what length scale disorder effects start to play a role. The crossover scale between
the Gaussian and the asymptotic critical behavior can be determined from the condition that the
renormalized dimensionless disorder coupling constant $V_R = V/\sqrt{tc_2^{3}} \approx 1$.
Now, the quantum phase transition occurs at $t=0$ which leads to $1/\tilde\Gamma_c=\ln({2\epsilon_F\tau})$,
with $\tilde\Gamma_c$ being the dimensionless critical coupling. Thus, the quantum phase
transition occurs at an exponentially small bare disorder strength which implies that $V_R \sim 1$
requires an exponentially large length scale.  Using the results above we find a Ginzburg type criterion:
\begin{equation}
\frac{|\tilde\Gamma - \tilde\Gamma_c|}{\tilde\Gamma_c} < \tilde\Gamma_c^3
        ~\exp \left[ -\frac{1}{\tilde\Gamma_c}\right]~.
\label{eq:ginzburg}
\end{equation}
Therefore, disorder effects become important only inside an exponentially narrow region
surrounding $\tilde\Gamma_c$. This asymptotic critical region is preempted by a wide Gaussian
crossover region (region below dashed dashed line on Fig. \ref{Fig:rg_flow}) with mean-field
critical behavior. For $p$-wave symmetry there are logarithmic correction to the power-law scaling
whose the $|{\bf q}|$ dependence reflects the underlying $p$-wave symmetry of the order parameter.

\section{Conclusions}
\label{Sec:Conclusions}

In this paper, we have studied the quantum phase transition from a dirty metal to an exotic
superconductor. Starting from a microscopic action of disordered interacting electrons, we have
derived the LGW theory for this quantum phase transition which proved to be equivalent (up to
logarithmic corrections in the Gaussian part in the case of $p$-wave pairing) to the
extensively studied LGW theory of the dirty itinerant antiferromagnetic transition. A
renormalization group analysis yielded runaway flow toward large disorder. As a result, the
asymptotic fate of the quantum phase transition is presently unknown. However, we could derive
a Ginzburg-type criterion for the importance of the disorder fluctuations. For weak bare
disorder, as is realized in many experimental systems, the true asymptotic behavior is observed
only exponentially close to the quantum critical point. It is preempted by a wide region with
mean-field behavior (and logarithmic corrections for $p$-wave pairing). In this last section we
analyze our results from a general mode-coupling point of view, and we also discuss
experiments.

In deriving the LGW functional, we have paid particular attention to the coupling between the
order parameter and additional soft modes. Mode-coupling induced singularities are indeed
present in all angular momentum channels, but they are increasingly suppressed for higher
angular momentum: In the static order parameter susceptibility the singular terms pick up an
extra power of $|{\bf q}|^{2L}$. This suppression can be understood as follows: In the presence
of non-magnetic quenched disorder, the dominant electronic soft modes are those that involve
fluctuations of the number density, spin density, or anomalous density in the zero angular
momentum channel \cite{bk_review} while the corresponding densities in higher angular momentum
channels are not soft. Since the different angular momentum modes are orthogonal at zero
wavenumber, the coupling between a finite angular momentum order parameter and the zero angular
momentum soft modes must involve powers of the wave number $|{\bf q}|$.
These arguments suggest a very general difference between the mode-coupling effects in clean
and in dirty electronic systems. While the only soft modes in the dirty case are in the zero
angular momentum channel, in clean systems, the charge, spin, and anomalous density
fluctuations in all angular momentum channels are soft (corresponding to an infinite number of
Fermi liquid parameters). Therefore, one expects the mode coupling singularities in a clean
system {\em not} to be suppressed by a higher order parameter angular momentum. This is known
to be true for the Cooper channel logarithmic singularities which do not pick up an extra
$|{\bf q}|^{2L}$ in clean electronic systems. A systematic investigation of this question will
be published elsewhere. \cite{unpublished}

The explicit calculations in this paper were for a superconducting quantum phase transition in a
paramagnetic system. We now discuss to what extend the results change if the transition happens in
a ferromagnetic system. Let us assume the magnetization being in $z$ direction. Obviously, not all
possible order parameter components are equivalent. Specifically, the symmetric triplet
$1/\sqrt{2}(\uparrow\downarrow+\downarrow\uparrow)$ (for $p$ or $f$-wave pairing) as well as the
singlet $1/\sqrt{2}(\uparrow\downarrow-\downarrow\uparrow)$ (for $s$ and $d$-wave pairing) are
suppressed because the exchange gap cuts off the Cooper-channel singularities. In contrast, for
equal spin pairing ($\uparrow\uparrow$ and $\downarrow\downarrow$), the leading behavior is the
same as discussed in Sections \ref{sec:Model} and \ref{sec:rg} of this paper.

Possible candidates for an experimental verification of our predictions  are the ferromagnetic
superconductors UGe$_2$ \cite{fm_sc_exp1} or ZrZn$_2$.\cite{fm_sc_exp2} For these systems, a
likely mechanism for superconductivity is $p$-wave triplet pairing mediated by magnetic
fluctuations due to the vicinity to a magnetic quantum critical point,\cite{us_fm_sc} although
this has not yet been established beyond doubt. In ZrZn$_2$ the vanishing of superconductivity
as a function of disorder has actually already been observed. \cite{fm_sc_exp2} A systematic
study of this transition would therefore be very interesting. \cite{qcp}

\begin{acknowledgments}
We thank D.\ Belitz, I.\ Herbut, T.\ R.\ Kirkpatrick, and S.\ Sessions for helpful discussions.
We acknowledge support from the German Research Foundation and from the University of Missouri
Research Board. Parts of this work have been performed at the University of Oxford (England),
at the Max-Planck-Institut f{\" u}r Physik komplexer Systeme, Dresden (Germany) and at the
Aspen Center for Physics.
\end{acknowledgments}

\appendix
\section{Two-point LGW vertex}
\label{appx:bubble}

In this Appendix we sketch the derivation of the expression (\ref{eqn:chi_2}) for the anomalous
density susceptibility $\chi^{(2)}$ in $3d$.  In a suitable parametrization and for $p$-wave
pairing, diagram Fig.\ \ref{Fig:chi2}a can be written as
\begin{equation}
D_{ij}^a({\mathbf q}, \omega )=T\sum_{\mathbf
k,\Omega}Y_1^i(\mathbf{\hat{e}}_\mathbf{k})Y_1^j(\mathbf{\hat{e}}_\mathbf{k})G(\mathbf k+\mathbf
q,\Omega+\omega)G(-\mathbf k,-\Omega),
\end{equation}
with $i$, $j$ being the order parameter component indices and ${\mathbf q}, \omega$ the
external momentum and frequency. $Y_L^M(\mathbf{\hat{e}}_{\mathbf k}) = Y_L^M(\theta,\phi)$ is
a spherical harmonic and Green's function $G$ is given by eq. (\ref{eq:Green_fun}).  A
straightforward calculation leads to:
\begin{equation}
\label{eq:appx2_Da}
D_{ij}^a=\frac{N_F}{3}\left[\log(2\epsilon_F\tau)-|\omega|\tau-\frac{1}{10}\ell^2|{\mathbf q}|^2\right]\delta_{ij}-\frac{N_F}{15}\ell^2q_iq_j.
\end{equation}
$\ell=k_F\tau/m$ is elastic mean free path and $N_F$ the density of states at the Fermi level.
Similarly, the diagram in  Fig.\ \ref{Fig:chi2}b is:
\begin{equation}
\label{eq:appx2_Db}
D^b_{ij}=\frac{N_F}{9}\ell^2q_iq_j\log(\omega\tau+\frac{\ell^2|{\mathbf q}|^2}{3}).
\end{equation}
Adding eqs.\ (\ref{eq:appx2_Da}) and (\ref{eq:appx2_Db}) completes derivation of eq.\
(\ref{eqn:chi_2}). For general angular momentum $L$, the analogous calculation shows that the
BCS logarithm in $D^b$ picks up an extra factor $|{\bf q}|^{2L}$ compared to the $s$-wave case.

\section{Interaction effects}
\label{appx:int}

\begin{figure}
\begin{center}
\includegraphics[scale=0.7]{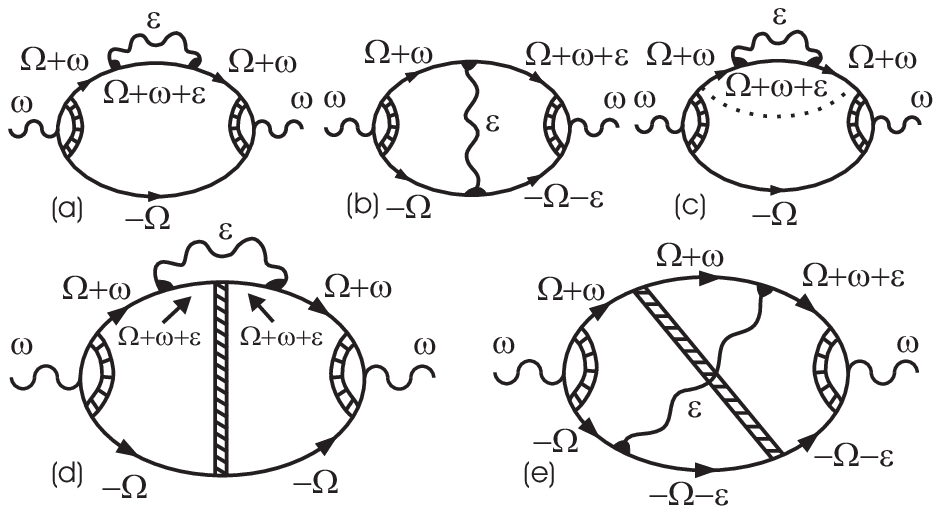}
\end{center}
\caption{Diagrams arising in the first order perturbation theory in interaction of the reference
ensemble $S_0$, and produce non-analytic, $|{\mathbf q}|^{2L} |{\mathbf q}|^{d-2}$, term, which is
a consequence of the mode-coupling effects.} \label{Fig:int}
\end{figure}

In this Appendix we analyze the leading corrections to $\chi^{(2)}$ due to the interactions
$S_{int}$ in the reference ensemble $S_0$. They can be understood as corrections to scaling at
the dirty Fermi liquid fixed point \cite{bk_review} and are particle-particle analogs of the
well known Altshuler-Aronov corrections to density of states and conductivity.\cite{Altshuler}
We first consider a paramagnetic reference ensemble. To first order in the interactions, the
relevant diagrams are those in Fig.\ \ref{Fig:int} and their counterparts with bare external
vertices. The wiggly line represents the interaction which is assumed to be short ranged and
can thus be approximated by a number $\Gamma_{\sigma\sigma'}$ (where $\sigma,\sigma'$ denote
the spin at the two ends of the interaction line).

Particular attention must be paid to the diagrams $a$, $b$ and $c$ with bare vertices. In these
diagrams the spherical harmonics in the two external vertices are not independent. Therefore,
their contributions can potentially produce stronger terms than $|{\bf q}|^{2L}|{\bf
q}|^{d-2}$. However, it turns out that these contributions do not produce any non-analytic
terms and only contribute to the regular terms. In the remaining contributions the angular
variables of the external vertices are independent. They can be analyzed along the lines
presented in Ref. \onlinecite{Altshuler:80}. After a straightforward calculation one finds that
the interaction corrections produce singularities of at most of the order $|{\bf q}|^{2L}|{\bf
q}|^{d-2}$ which means they are suppressed by a factor $|\mathbf q|^{2L}$ compared to the zero
angular momentum case.\cite{kb_dirty_fm}

The above conclusion is easily generalized to ferromagnetic reference ensembles: If the
magnetization is in $z$-direction, the $\uparrow\uparrow$ and $\downarrow\downarrow$ components
of the order parameter have the same type of singularity as discussed above, while the leading
singularities in the $\uparrow\downarrow$ components are cut-off by the exchange gap.

\section{4-point vertices}
\label{appx:football}

In this Appendix  we present details of calculation of four-point susceptibility $\chi^{(4)}$
in $3d$, eqs. (\ref{eq:chi_4diag}) and (\ref{eq:chi_4off}). We start with replica diagonal
part, $\chi^{(4)}_{diag}$ which is calculated from the diagrams shown on the Fig.\
\ref{Fig:chi4d}. The most singular contributions are produced if the frequencies structure
permits all four external vertices to be renormalized by an active (retarded-advanced) ladder
(4-ladder diagrams). A direct calculation of the first diagram (Fig.\ \ref{Fig:chi4d}a) leads
to:
\begin{eqnarray}
\label{eq:appx_3_Da2}
D_a=\frac{2\ell^5\tau^2|{\mathbf q}|^4\cos^4(\alpha)}{81\pi}\sum_{\Omega}\frac{\Theta(\Omega(\Omega+\omega))}{(|2\Omega+\omega|\tau+\frac{\ell^2|{\mathbf q}|^2}{3})^4}\times\nonumber\\
\times(1-9|2\Omega+\omega|\tau-\ell^2|{\mathbf q}|^2).
\end{eqnarray}
An analogous calculation for diagram \ref{Fig:chi4d}b) gives (the frequency constraint requires the
extra impurity line to act as a single impurity line rather than an active ladder):
\begin{eqnarray}
\label{eq:appx_3_Db2}
D_b=-\frac{\ell^5\tau^2|{\mathbf q}|^4\cos^4(\alpha)}{81\pi}\sum_{\Omega}\frac{\Theta(\Omega(\Omega+\omega))}{(|2\Omega+\omega|\tau+\frac{\ell^2|{\mathbf q}|^2}{3})^4}\times\nonumber\\
\times(1-12|2\Omega+\omega|\tau-2\ell^2|{\mathbf q}|^2).
\end{eqnarray}
Each of these two diagrams individually has an divergence  $\sim 1/|{\bf q}|^2$. However,
because the relative combinational factor of $D_b$ is $2$, the divergent contributions cancel,
rendering the final value for $\chi^{(4)}_{diag}$ finite (Eq.\ (\ref{eq:chi_4diag})).

\begin{figure}[b]
\includegraphics[scale=0.6]{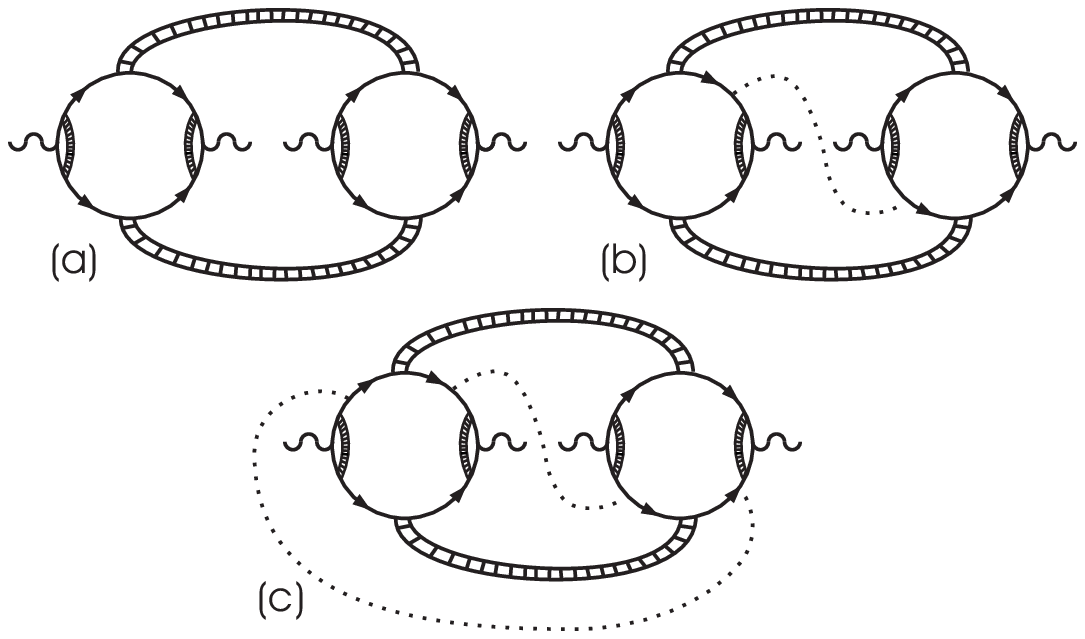}%
\caption{Diagrams contributing to the $\chi^{(4)}_{off}$. All four vertices are renormalized, with
two extra active particle-particle ladders connecting two fermionic loops. The zeroth order terms
in small $\mathbf q$ and $\omega$ of these diagrams cancel.} \label{Fig:goggle}
\end{figure}
Similar cancellations among individually diverging diagrams take place in the replica off-diagonal
case, with the strongest singularities coming from diagrams with the largest number of active
ladders. A set of such diagrams is shown on Fig.\ \ref{Fig:goggle}. Here, at most six  ladders can
be active, leading to an infrared singularity in each of the diagrams of the form $1/|{\bf q}|$.
Similar calculations to the ones carried out above reveal that the singular contributions of
diagrams (a), (b), and (c) canceled each other. The remaining contribution is finite and can be
estimated from the simple diagram Fig.\ \ref{Fig:goggle_bare}.

We emphasize once more that all the results for the singularities in the anomalous density
correlation functions in the Appendices A to D have been obtained in low-order perturbation
theory. Within perturbation theory one cannot prove that the terms obtained indeed represent the
leading singularities to all orders. We are nonetheless confident that we indeed identified the
leading terms, because in the $s$-wave case we reproduce the known rigorous results from $Q$-field
theory.\cite{kb_scqpt_97,bk_review}


\begin{thebibliography}{99}
\bibitem{afm} For finite-$q$ order parameters, the coupling to fermionic
   modes is generically weaker, and Hertz' scheme works, at least in 3d.
   \cite{AbanovChubukov}
\bibitem{divvert}Note that the 6th and higher order terms in the LGW functional are
   renormalization-group irrelevant even though their coefficients are singular.
\bibitem{qcp} Directly at the magnetic quantum critical point our theory may not be applicable because of strong
      non-Fermi liquid effects and the interplay between spin-fluctuations and pairing
      \cite{chubukov_1st}. However, away from the magnetic quantum critical point the
      low-temperature normal state is a Fermi liquid, and our theory should apply.
\end{thebibliography}
\end{document}